\documentstyle[12pt]{article}
\begin{document}

\overfullrule 0 mm
\language 0
\centerline { \bf{ FAILURE OF LORENTZ-DIRAC APPROACH}}
\centerline { \bf{ TO RADIATION REACTION}}
\centerline { \bf{ FOR EXTREMELY LARGE
VELOCITIES ?}} \vskip 0.5 cm \centerline {\bf{ Alexander A.  Vlasov}}
\vskip 0.3 cm
\centerline {{  High Energy and Quantum
Theory}} \centerline {{  Department of Physics}} \centerline {{
Moscow State University}} \centerline {{  Moscow, 119899}}
\centerline {{  Russia}}
\vskip 0.3 cm
{\it
 On the model of moving rigid charged body is shown that the
Lorentz-Dirac approach to radiation reaction in classical
electrodynamics does not work for trajectories of body close to the
light cone.}

 03.50.De \vskip 0.3 cm

 Since the famous Dirac's paper on relativistic radiation reaction
in classical electrodynamics, many textbooks and research articles
were published on that theme. Among them are [1-11], where one can
find the discussion of the related problems: mass renormalization and
its nonuniqueness, runaway solutions and the use of the advanced
interaction.( These problems of radiation reaction one can find
also in other classical field theories - scalar field theory,
gravitational theory, etc.). For recent review of the problems see
[12].

Here we present the new puzzle connected with the Lorentz-Dirac
approach.

In the literature  the derivations of relativistic form of
radiation force are made under the "hidden" assumption that all the
expansions in powers used near the particle trajectory are valid for
the whole range of values of particle velocity, in
particular, arbitrary close to that of the light.

But it is easy to see that this is not true in general.

 For example let's  consider the model of extended
particle  with charge $Q$, density $\rho$ and
density current $\vec j= \vec v \cdot \rho$.

 Then the force, acting on this particle from self electromagnetic
 field is
 $$\vec F_{self} = \int dv  \rho \left(\vec
E +[\vec v, \vec H]/c \right) \eqno(1)$$

If the particle is compact enough, one can use in (1) for
  electromagnetic fields $\vec E$ and $\vec H$ the explicit
  Lienard-Wiechert expressions:
$$\vec E(t,\vec r)= Q(1-{v^2 \over c^2}){\vec L - \vec{ v} L/c \over
(L-\vec {v} \vec {L} /c)^3} +Q { [\vec {L}, [\vec {L} - \vec {v} L/c
,\dot {\vec {v}} /c]]\over c (L-\vec v \vec {L} /c)^3},$$ $$ \vec
H=[\vec L,\vec E]/L \eqno(2)$$ here $$\vec L =\vec r- \vec
R(t_{ret})$$ $$t_{ret}= t -|\vec r- \vec R(t_{ret})|/c \eqno(3)$$.
 It is convenient to rewrite (3) in the form of the equation:  $$\vec
R(t_{ret})=\vec R(t-|\vec r- \vec R(t_{ret})|/c) \eqno (4)$$ Let the
trajectory of the particle "center" be $\vec R(t)$.  Then for compact
 body the value of $\vec r$ is close to $\vec R(t)$:  $$\vec r= \vec
 R(t) +\vec \mu \eqno(5)$$ with $|\vec \mu| \sim a \ll R$ and $a$ -
 is the size of the body.  With new vector $\vec \nu$ $$\vec
R(t_{ret})= \vec R(t) +\vec \nu \eqno(6),$$ the equation (4) for
$\vec \nu=\vec \nu(t,\vec \mu)$ takes the form $$\vec R(t) +\vec \nu
=\vec R(t-|\vec \nu -\vec \mu|/c) \eqno (7)$$ One solution of (7) is
obvious:

for  $\vec \mu=0$ one has $\vec \nu=0$

(i.e. the body is point-like)

If the body is stretched ($\vec \mu \not= 0$) then to find
solution of (7) one can try to use the expansion in powers of
$\vec \mu$.

And this is what is done in literature, while extracting from (1)
the expression for the radiation force for small object ($a\to 0$).

 But this method failures for extremely large velocity $|v|\to c$.

 Indeed  let's take the following form of trajectory:
$$\vec R =(0,0,R(t))$$
$$R(t)= ct+b/t,\ \ b=const, t\to \infty \eqno(8)$$
(One can consider (8) as the first terms of expansion in
powers of $O(b/(ct^2))$ of the  hyperbolic motion of a particle:
$R= (c/A)\sqrt{1+(At)^2},\ \ b={c\over 2A^2}$)

If the body is stretched in $z$-dimension, then $\vec \mu
=(0,0,\mu)$ and thus the equation (7) for  $\nu=  \nu(t,\mu)$
takes the form:  $$ R(t) + \nu = R(t-|\nu - \mu|/c) \eqno (9)$$

If to search the solution of (9) with the following property:
$\nu \to 0$ with $\mu \to 0$, then for (8) it is
$$\nu=\mu- {c t^2 \mu \over b+t\mu}\eqno(10)$$
Thus we see that if $\mu=0$ then $\nu=0$, and for $b \gg
t\mu $ one can expand the solution (10) in powers of $t\mu$. On this
way the standard relativistic Lorentz-Dirac expression for the
radiation force is derived.

But if $t\mu \gg b$ ( the moments of
time are great enough compared with the inverse "size" of the
extended body), then  this expansion does not work (the asymptotic
value for $\nu$ ($t\to \infty$) is $-ct$) and thus the Lorentz-Dirac
result is not valid.

Let's examine this idea in details.

Consider the model of  the rigid charged sphere
with radius $a$ moving with velocity $\vec v(t)=d\vec R(t) /dt$. The
total charge of the sphere is $Q$ with  density $\rho(t,\vec
r)={Q\over 4\pi a^2}\delta(|\vec r- \vec R|-a)$.   (We are aware that
physically this model is not realized - in the relativity theory
there are no absolutely rigid bodies, so we consider this model only
as convenient mathematical tool). Then the Maxwell equations give the
following expression for the retarded  electric field of this
sphere:

$$\vec E(t,\vec r) = {Qc\over 2a}\int\limits_{t_1}^{t_2}dt'{\vec
N'\over |\vec r- \vec R'|^2} +{Q\over 2a}\left[ {k\vec N_2 -\vec
v_2/c \over|\vec r- \vec R_2|(1-k\vec N_2 \vec v_2/c)}-{\vec N_1
-\vec v_1/c \over|\vec r- \vec R_1|(1-\vec N_1 \vec v_1/c)} \right]
\eqno(9)$$
here
$$t_2=t-||\vec r- \vec R(t_2)|-a|/c$$
$$t_1=t-||\vec r- \vec R(t_1)|+a|/c$$
$$\vec R' = \vec R(t'),\ \ \vec R_a=\vec R(t_a),\ \ \vec v_a=\vec
v(t_a), \ \ a=1,2,$$
$$ k=(|\vec r- \vec R(t_2)|-a)/||\vec r- \vec
R(t_2)|-a|$$.

For simplicity let the sphere move as a rigid body along the
$z$-axis.

 To find the total force $\vec F_{self}$, acting on the moving sphere
from its electromagnetic field, one must integrate the expression
(9) over the sphere (due to the motion on a straight line the Lorentz
force is absent).  This gives the following expression for the $z$-
component of the force:  $$F_{z, self} = {Q^2c\over
4a}\int\limits_{-1}^{+1}dx\int\limits_{t_1}^{t}dt'{ax+R-R'\over
 [a^2+2ax(R-R')+(R-R')^2]^{3/2}}$$
$$ +{Q^2\over
4a}\int\limits_{-1}^{+1}dx
{x-v/c \over a(1-xv/c)}  $$
$$-{Q^2\over
4a}\int\limits_{-1}^{+1}dx
{\vec N_1 -\vec v_1/c \over [a^2+2ax(R-R_1)+(R-R_1)^2]^{1/2}
(1-\vec N_1 \vec v_1/c)}  \eqno(10)$$
here
$$t_1=t- [a^2+2ax(R-R_1)+(R-R_1)^2]^{1/2}/c-a/c,$$
$$N_1={ax+R-R_1\over
 [a^2+2ax(R-R_1)+(R-R_1)^2]^{1/2}}  $$

Then if to assume the trajectory of the sphere in the form  (8), then
two last terms in (10) can be calculated explicitly:

$${Q^2\over 2a^2} \left(-{2c \over v}+(1-{c^2 \over v^2})\ln
{1-v/c \over 1+v/c} \right)$$
$$+{Q^2\over 4a^2}\left( -2\ln{\Delta_{+}\over \Delta_{-}}
+\ln{\tau_{+}-a \over \tau_{-}-a}\left[1-{(1-{a\over ct})^2 \over
(1- {b\over ct^2}-{a\over ct})^2} \right] \right)$$
 $$+{Q^2\over 4a^2}\left(+\ln{\tau_{+}-ct(1- {b\over ct^2})
\over \tau_{-}-ct(1- {b\over ct^2})}\left[1+{(1-{a\over ct})^2 \over (1-
{b\over ct^2}-{a\over ct})^2} -{2\over (1- {b\over ct^2})^2 }\right]
\right)$$
$$+{Q^2\over 4a^2}\left(+{2\over (1- {b\over ct^2})^2 } \ln{\tau_{+} 
\over
\tau_{-}}\right)$$
$$+{Q^2\over 4a^2}\left( {b^2 \over ct^3}\left[ {2\over 1- {b\over 
ct^2}}
-{1\over 1- {b\over ct^2}-{a\over ct}}\right]\left[{1\over \tau_{+}-
ct(1-
{b\over ct^2})} -{1\over \tau_{-}-ct(1- {b\over ct^2})}\right]\right)
\eqno(11)$$
with
$$\tau_{\pm}=ct-ct_{\pm}$$
$$\Delta_{\pm}=R(t)-R(t_{\pm})$$
and $t_{\pm}$ -  the solutions of the equations:
$$t_{\pm}= t -a/c -[a^2 \pm
2a\Delta_{\pm}+(\Delta_{\pm})^2]^{1/2}/c$$
Under the assumption that the moments of time are large enough
compared with the size of the body $a$,
$$t \gg b/a \gg a/c \eqno (12)$$
we can calculate approximately the first term in (10):
$$+{Q^2\over 4a}\left( {2\over a} +{2b\over ct^2 a}\ln{b \over
ct^2}-{2\over ct}+O({b\over
c^2t^3})+O({a^2\over bc^2t})\right)\eqno(13)$$
The use of (12) in (11) gives
$$+{Q^2\over 4a}\left( -{2\over a} -{2b\over ct^2 a}\ln{b \over
ct^2}- {2b \over ct^2a}\ln{a \over ct}-{2\over ct}+O({b\over
c^2t^3})\right)\eqno(14)$$
The sum of (13) and (14) yields the following result for the total
force:  $$F_{z,self}= {Q^2
 \over 4a}\left[ -{4\over ct}- {2b \over ct^2a}\ln{a \over ct}+ O({b
 \over c^2t^3}) +O({ a^2 \over b c^2t}) \right]\eqno (15)$$

Thus we see that the result (15) is not what  one can suspect
from the standard Lorentz-Dirac approach in the limit $a\to 0$, when
$$F= -m_{em}w +...\eqno(16)$$ with $m_{em}$ - the so called
electromagnetic mass of the body, $m_{em} \sim {Q^2 \over a}$ and $w$
- is the  acceleration, $W= {d \over dt}{v \over
\sqrt{1-(v/c)^2}  } \approx {c^{3/2}\over (2b)^{1/2}}(1+{b \over
ct^2})$ for the eq.  of motion in the form (8).

The force, acting on the body due to our result (15) is smaller then
(16) and for $t\to \infty$, $a$ -small, but $a\not=0$, tends to zero.

Due to (8) the relativistic acceleration $W={d^2 R \over ds^2}$ is
$W\approx {t\over 2b}$. So we can formulate our conclusion in the
form:

{\bf for particle with infinitesimaly small size $a$ moving along
trajectory $R(t)$ the Lorentz-Dirac result for radiation force meets
with failure for those moments of time for trajectories close to
light cone, when the following inequiality becomes valid:}
$$1\leq Wa$$.

 \vskip 0.5 cm \centerline {\bf{REFERENCES}}

  \begin{enumerate}
  \item
  F.    Rohrlich, {\it Classical Charged Particles}, Addison-Wesley,
  Reading, Mass., 1965.
\item
D.Ivanenko, A.Sokolov,  {\it Classical field theory} (in
russian), GITTL, Moscow, 1949.
A.Sokolov, I.Ternov, {\it Syncrotron Radiation}, Pergamon Press,
    NY, 1968. A.Sokolov, I.Ternov, {\it Radiation from Relativistic
Electron}, AIP, NY, 1986.
\item
G.Plass, Rev.Mod.Phys., 33, 37(1961).
\item S.Parrott, {\it
Relativistic Electrodynamics and Differential Geometry},
 Springer-Verlag, NY, 1987.

 \item C.Teitelboim, Phys.Rev., D1, 1572 (1970); D2,
1763 (1970).
Teitelboim, D.Villaroel, Ch. G. van Weert Riv. Nuovo Cim. 3, 1
(1980). R.Tabensky, Phys.Rev., D13, 267 (1976).
 \item  E.Glass, J.Huschilt and G.Szamosi, Am.J.Phys.,
52, 445 (1984).
 \item S.Parrott, Found.Phys., 23,
1093 (1993).
\item W.Troost et al.,  preprint hep-th/9602066.
\item Alexander A.Vlasov, preprints hep-th/9702177;
hep-th/9703001, hep-th/9704072.
  \item J.Kasher, Phys.Rev., D14,
939 (1976).
J.Huschilt, W.E.Baylis, Phys.Rev., D17,
985 (1978).
 \item S. de Groot, L.Suttorp {\it Foundations of
Electrodynamics}, North-Holland, Amsterdam, 1972
\item  Alexander A.Vlasov, preprint hep-th/9707006.

\end{enumerate}

 \end{document}